\documentclass[12pt]{article}

\def\ben{\begin{equation}}
\def\een{\end{equation}}

  \let\n=\nu

\let\C=\Chi

\def\nn{\nonumber} \def\bd{\begin{document}} \def\ed{\end{document}}
\def\ds{\documentstyle} \let\fr=\frac \let\bl=\bigl \let\br=\bigr
\let\Br=\Bigr \let\Bl=\Bigl
\let\bm=\bibitem
\let\na=\nabla
\let\pa=\partial \let\ov=\overline
\newcommand{\be}{\begin{equation}}
\newcommand{\ee}{\end{equation}}
\def\ba{\begin{array}}
\def\ea{\end{array}}
\def\ft#1#2{{\textstyle{{\scriptstyle #1}\over {\scriptstyle #2}}}}
\def\fft#1#2{{#1 \over #2}}
\def\del{\partial}
\def\vp{\varphi}
\def\sst#1{{\scriptscriptstyle #1}}
\def\oneone{\rlap 1\mkern4mu{\rm l}}
\def\td{\tilde}
\def\wtd{\widetilde}
\def\ie{\rm i.e.\ }
\def\dalemb#1#2{{\vbox{\hrule height .#2pt
        \hbox{\vrule width.#2pt height#1pt \kern#1pt
                \vrule width.#2pt}
        \hrule height.#2pt}}}
\def\square{\mathord{\dalemb{6.8}{7}\hbox{\hskip1pt}}}
\newcommand{\ho}[1]{$\, ^{#1}$}
\newcommand{\hoch}[1]{$\, ^{#1}$}
\newcommand{\bea}{\begin{eqnarray}}
\newcommand{\eea}{\end{eqnarray}}
\newcommand{\ra}{\rightarrow}
\newcommand{\lra}{\longrightarrow}
\newcommand{\Lra}{\Leftrightarrow}
\newcommand{\ap}{\alpha^\prime}
\newcommand{\bp}{\tilde \beta^\prime}
\newcommand{\tr}{{\rm tr} }
\newcommand{\Tr}{{\rm Tr} }
\def\0{{\sst{(0)}}}
\def\1{{\sst{(1)}}}
\def\2{{\sst{(2)}}}
\def\3{{\sst{(3)}}}
\def\4{{\sst{(4)}}}
\def\5{{\sst{(5)}}}
\def\6{{\sst{(6)}}}
\def\7{{\sst{(7)}}}
\def\8{{\sst{(8)}}}
\def\n{{\sst{(n)}}}
\def\cA{{{\cal A}}}
\def\cB{{{\cal B}}}
\def\cF{{{\cal F}}}
\def\tV{\widetilde V}
\def\tW{\widetilde W}
\def\tH{\widetilde H}
\def\tE{\widetilde E}
\def\tF{\widetilde F}
\def\tA{\widetilde A}
\def\im{{{\rm i}}}
\def\tY{{{\wtd Y}}}
\def\ep{{\epsilon}}
\def\vep{{\varepsilon}}
\def\R{\rlap{\rm I}\mkern3mu{\rm R}}
\def\bD{{{\bar D}}}

\def\R{\rlap{\rm I}\mkern3mu{\rm R}}
\def\bD{{{\bar D}}}
\def\R{{{\Bbb R}}}
\def\C{{{\Bbb C}}}
\def\H{{{\Bbb H}}}
\def\CP{{{\Bbb C}{\Bbb P}}}
\def\RP{{{\Bbb R}{\Bbb P}}}
\def\Z{{{\Bbb Z}}}
\def\bA{{{\Bbb A}}}
\def\bB{{{\Bbb B}}}
\def\bC{{{\Bbb C}}}
\def\bD{{{\Bbb D}}}
\def\bE{{{\Bbb E}}}
\def\bZ{{{\Bbb Z}}}
\def\Re{{{\frak{Re}}}}
\def\Im{{{\frak{Im}}}}
\def\cosec{{\,\hbox{cosec}\,}}
\def\Gm{{\Gamma_{\!\! -}}}
\def\Gp{{\Gamma_{\!\! +}}}
\def\stan{{standard }}
\def\nonstan{{supernumerary }}

\newcommand{\tamphys}{\it Center for Theoretical Physics,
Texas A\&M University, College Station, TX 77843}

\newcommand{\upenn}{\it Department of Physics and Astronomy,\\ University
of Pennsylvania, Philadelphia, PA 19104}

\newcommand{\brussels}{\it Physique Th\'eorique et Math\'ematique,
Universit\'e Libre de Bruxelles,\\ Campus Plaine C.P. 231, B-1050
Bruxelles, Belgium}

\newcommand{\auth}{J. Kerimo and H. L\"u\hoch{1}}

\begin{document}
\begin{flushright}

MIFP-03-16\\
{\bf hep-th/0307222}\\
July\  2003
\end{flushright}

\vspace{10pt}

\begin{center}

{\large {\bf New $D=6$, ${\cal N}=(1,1)$ Gauged Supergravity with
Supersymmetric (Minkowski)$_4\times S^2$ Vacuum}}

\vspace{20pt}
\auth

\vspace{20pt} {\it George P. and Cynthia W. Mitchell
Institute for Fundamental Physics,\\ Texas A\& M University,
College Station, TX 77843-4242, USA}

%\vspace{10pt} {\hoch{\dagger}\brussels}

\vspace{40pt}

\underline{ABSTRACT}
\end{center}

      We obtain a new gauged $D=6$, ${\cal N}=(1,1)$ pure supergravity
by a generalised consistent Kaluza-Klein reduction of M-theory on
K3$\times R$.  The reduction requires a conspiratory gauging of both
the Cremmer-Julia type global (rigid) symmetry and the homogeneous
rescaling symmetry of the supergravity equations of motion.  The
gauged supergravity is different from the Romans $D=6$ gauged
supergravity in that the four vector fields in our new theory are all
abelian.  We show that it admits a supersymmetric (Minkowski)$_4\times
S^2$ vacuum, which can be lifted to $D=11$ where it becomes the
near-horizon geometry of two intersecting M5-branes wrapping on a
supersymmetric two-cycle of K3.

{\vfill\leftline{}\vfill \vskip 10pt \footnoterule {\footnotesize
\hoch{1} Research supported in part by DOE grant
DE-FG03-95ER40917.}}

\pagebreak
\setcounter{page}{1}

%\tableofcontents
%\addtocontents{toc}{\protect\setcounter{tocdepth}{2}}
\newpage

\section{Introduction}

     It has long been known that the Salam-Sezgin model of $D=6$,
${\cal N}=(1,0)$ Einstein-Maxwell gauged supergravity admits a
supersymmetric (Minkowski)$_4\times S^2$ vacuum with the chiral ${\cal
N}=1$ four-dimensional supersymmetry \cite{szmodel}. It was recently
shown that the six-dimensional theory admits a fully consistent
dimensional reduction on the 2-sphere \cite{gp}, giving rise to a
massless sector that comprises a supergravity multiplet, an $SU(2)$
Yang-Mills multiplet and a scalar multiplet, with a vanishing
cosmological constant.  The supersymmetry and the absence of the
cosmological constant make it an interesting model for phenomenology
\cite{abpq1,abpq2,gp}.

      However, the higher-dimensional origin of the six-dimensional
theory remains elusive.  It is a particular case of a more general class
of supergravities constructed in \cite{ns}, but it cannot be obtained
from the truncation of the Romans $D=6$ ${\cal N}=(1,1)$ gauged
supergravity \cite{romans}.  A potential difficulty is associated
with the fact that the scalar potential is positive definite, which
rules out the possibility of a sphere reduction, which is a typical
origin of gauged supergravities.  Recently, it was proposed that the
six-dimensional theory should be viewed as an effective theory on the
boundary of AdS$_7$; however, the reduction is not consistent in the
bulk \cite{sorin1,sorin2}.

       In this paper we obtain a new $D=6$, ${\cal N}=(1,1)$ gauged
supergravity, which has also a (Minkowski)$_4\times S^2$ vacuum, from
a consistent {\it generalised} Kaluza-Klein reduction of $D=11$
supergravity on K3$\times R$.  The dimensional reduction comprises two
steps.  The first step is to compactify M-theory on a K3 manifold,
which gives rise to, as the low-energy effective theory, the $D=7$,
${\cal N}=1$ supergravity coupled to matter.\footnote{The consistency
of the reduction of M-theory on K3 is questionable. It is consistent
however if one restricts to the pure supergravity multiplet, which
does not turn on the scalars parameterising the K3.}  We shall
concentrate on the pure supergravity multiplet, which consists of the
metric, the dilaton, one rank-2 antisymmetric tensor, three vectors,
and their fermonic counterparts.  This theory has a Cremmer-Julia type
global (rigid) symmetry which shifts the dilaton by a constant.  The
Lagrangian is invariant provided that appropriate constant scalings of
the tensor and vector fields are performed.  The $D=7$ theory has also
a homogeneous scaling symmetry, with the following transformation rule
%%%%
\be
ds^2\rightarrow \Omega^2\, ds^2\,,\qquad
A_\n \rightarrow \Omega^n\, A_\n\,,\qquad
\Psi_{\sst{M}}\rightarrow \Omega^{\ft12}\,\Psi_{\sst{M}}\,,
\ee
%%%%
where $A_\n$ is an $n$-form gauge potential and $\Psi_M$ is the
gravitino, with $M$ being the curved index.  This transformation has
the effect of scaling the whole Lagrangian, and therefore it is not a
symmetry of the action, but only of the equations of motion.  It was
shown in \cite{lav} that both symmetries can be gauged in a
Kaluza-Klein circle reduction, where the constant parameters of the
global symmetry transformation rules are taken to be linearly
dependent on the reduction circle coordinate.  It was furthermore
shown that this procedure of reduction is consistent; it is called
``generalised Kaluza-Klein reduction'' in \cite{lav}.

      The motivation for us to consider such a generalised reduction
ansatz is that the gauging of the homogeneous rescaling symmetry has
the effect of generating a positive cosmological constant \cite{lav}.
Thus, in the second step, we apply such a generalised reduction on the
minimal $D=7$ supergravity and obtain a new $D=6$ ${\cal N}=(1,1)$
gauged supergravity coupled to a vector multiplet.  We find that the
matter vector multiplet can remarkably be truncated out by a certain
choice of the two cosmological parameters associated with the gauging
of the two symmetries, giving rise to a pure $D=6$, ${\cal N}=(1,1)$
gauged supergravity.

          The new theory is different from the Romans $D=6$, ${\cal
N}=(1,1)$ gauged supergravity in that the four vectors in our new
theory are all abelian instead of being $SU(2)\times U(1)$ Yang-Mills
fields.  An interesting feature of the $D=6$, ${\cal N}=(1,1)$ theory
that we are going to present is that it also admits a supersymmetric
(Minkowski)$_4\times S^2$ vacuum, with the ${\cal N}=2$
four-dimensional supersymmetry.  Owing to the consistency of our
reductions, we can lift the solution back to $D=11$, where it becomes
the near-horizon geometry of two intersecting M5-branes wrapping on a
supersymmetric two-cycle of K3.  The solution of the two intersecting
M5-branes preserves $\ft14$ of the maximal supersymmetry.

         Since the compactification of M-theory on K3 is well
understood, we consider in section 2 the second step of the reduction,
namely the generalised Kaluza-Klein reduction of the minimal
supergravity in $D=7$ on $S^1$.  We obtain the new ${\cal N}=(1,1)$
gauged supergravity in six dimensions coupled to a vector multiplet.
We show that by setting equal the two arbitrary mass parameters, the
vector multiplet can consistently be truncated out, giving rise to the
pure $D=6$, ${\cal N}=(1,1)$ gauged supergravity.  In addition, we
show that a further truncation in the bosonic equations of motion to a
subsector is possible, whose Lagrangian turns out to be identical to
the bosonic sector of the Salam-Sezgin model.  However, this
truncation is not supersymmetric.  In section 3, we obtain the
supersymmetric (Minkowski)$_4\times S^2$ vacuum solution.  We first
lift the solution back to $D=7$, where it becomes the near-horizon
geometry of the 3-brane in $D=7$.  It can then further be lifted back
to $D=11$, and the solution becomes the near horizon geometry of two
intersecting M5-branes wrapping a supersymmetric two-cycle of the K3.
We provide further comments and conclusions in section 4.

\section{New gauged ${\cal N}=(1,1)$ supergravity from $D=7$}

      Minimal supergravity in seven dimensions consists of the metric,
a dilaton, a 2-form antisymmetric tensor and three vectors, together
with their fermionic counterparts.  The Lagrangian for the bosonic
sector is given by
%%%%%
\be
\hat e^{-1}{\cal L} = \hat R - \ft12 (\del\hat \phi)^2 - \ft{1}{12}
e^{\ft4{\sqrt{10}}\,\hat \phi}\, \hat G_\3^2 -
\ft14\, e^{\ft2{\sqrt{10}}\,\hat\phi}\,
(\hat F_\2^i)^2\,,\label{d7lag1}
\ee
%%%%%%
where $\hat G_\3=d\hat B_\2 + \ft12 \hat F_\2^i\wedge \hat A_\1^i$,
$\hat F_\2^i=d\hat A_\1^i$ and $i=1,2,3$.  The theory possesses the
following global (rigid) symmetry
%%%%
\bea
&&\hat \phi\rightarrow \hat \phi + \sqrt{10}\,\lambda_1
\,,\qquad d\hat s^2\rightarrow e^{2\lambda_2}\, d\hat s^2\,,\nn\\
&& \hat B_\2\rightarrow e^{-2\lambda_1 + 2\lambda_2}\, \hat B_\2\,,
\qquad \hat A_\1^i \rightarrow e^{-\lambda_1 + \lambda_2}\,
\hat A_\1^i\,.
\eea
%%%%
The transformation associated with $\lambda_1$ leaves the Lagrangian
invariant, and therefore describes a symmetry of the Lagrangian. On
the other hand, the transformation associated with $\lambda_2$ scales
the whole Lagrangian, and so it is not a symmetry of the Lagrangian,
or even the action, but a symmetry of the equations of motion.

        It was shown in \cite{lav} that these symmetries can be gauged
in the Kaluza-Klein circle reduction, since we can let $\lambda_i$
depend linearly on the $S^1$ coordinate $z$.  Following \cite{lav}, we
consider the following generalised $S^1$ reduction ansatz
%%%%
\bea
d\hat s_7^2 &=& e^{2m_2\, z}\, \Big(e^{2\alpha\,\varphi}\, ds_6^2 +
e^{2\beta\,\varphi}\, (dz + \cA_\1)^2\Big)\,,\nn\\
\hat B_\2 &=& e^{2(m_2-m_1)\,z}\, (B_\2 +
B_\1\wedge dz )\,, \nn\\
\hat A_\1^i &=& e^{(m_2-m_1)\,z}\, (A_\1^i +
\chi^i\, dz)\,,\nn\\
\hat \phi &=& \phi + \sqrt{10}\, m_1\, z\,,\label{d7red1}
\eea
%%%%%% 
where $\alpha^2=\ft{1}{40}$ and $\beta=-4\alpha$.  When $m_1=0=m_2$,
the ansatz becomes the one for the standard $S^1$ Kaluza-Klein
reduction.  We shall next present the detailed reduction and show that
the generalised reduction ansatz (\ref{d7red1}) is consistent with the
equations of motion of the $D=7$ minimal supergravity.  The resulting
$D=6$ theory is somewhat complicated, involving mass parameters $m_1$
and $m_2$.  One reason for this complication is the occurrence of the
prefactors for $\hat B_\2$ and $\hat A_\1^i$ in the reduction ansatz
(\ref{d7red1}).  However, the prefactors vanish if we set
$m_1=m_2\equiv m$.  In this case, the system becomes much simpler.  In
particular, it is possible to consistently truncate out the axionic
fields $\chi^i$ and set equal the Kaluza-Klein and winding vectors
$B_\1$ and $\cA_\1$.  Furthermore, one combination of the dilaton and
$\varphi$ can be set to zero.  The resulting theory is a new $D=6$,
${\cal N}=(1,1)$ pure gauged supergravity.

     First, we find that the field strengths $\hat G_\3$ and $\hat
F_\2^i$ can be expressed as
%%%%
\bea
\hat G_\3 &=& e^{2(m_2-m_1)\,z}\,
\Big( G_\3 + G_\2\wedge (dz + \cA_\1) \Big)\,,\nn\\
\hat F_\2^i &=& e^{(m_2-m_1)\, z}\,
\Big( F_\2^i + L_\1^i\wedge (dz + \cA_\1)\Big)\,,
\eea
%%%%%
where
%%%%
\bea
G_\3\!\!\! &\equiv&\!\!\! dB_\2 + \ft12 F_\2^i\wedge A_\1^i -
dB_\1\wedge \cA_\1 - 2(m_2-m_1)\, B_\2\wedge \cA_\1 -
\ft12\chi^i\, F_\2^i\wedge \cA_\1\,,\nn\\
G_\2\!\!\!&\equiv&\!\!\! dB_\1 + \ft12 \chi^i\, F_\2^i -
\ft12 L_\1^i\wedge A_\1^i + \ft12\chi^i\, L_\1^i\wedge \cA_\1 +
2(m_2-m_1)\, B_\2\,,\nn\\
F_\2^i&\equiv& dA_\1^i - d\chi^i\wedge\cA_\1 +
(m_2-m_1)\, A_\1^i\wedge A_\1\,,\nn\\
L_\1^i&\equiv& d\chi^i - (m_2-m_1)\, A_\1^i\,.
\eea
%%%%%%
Thus, we see that for general values of $m_1$ and $m_2$, the vector
fields $A_\1^i$ and the tensor field $B_\2$ become massive, eating the
axions $\chi^i$ and the winding vector $B_\1$ respectively.  Their
masses are proportional to $|m_2-m_1|$, and therefore vanish when
$m_1=m_2$.

          Substituting the reduction ansatz into the equations of
motion in $D=7$, we find that they can all be satisfied provided that
the lower dimensional fields satisfy the following equations of motion
%%%%%
\bea
&&\nabla^\theta\Big(e^{a\,\phi -4\alpha\,\varphi}\, G_{\mu\nu\theta}
\Big) = (3m_2 + 2m_1)\, \Big(
e^{a\,\phi - 4\alpha \varphi}\, G_{\mu\nu\theta}\, \cA^\theta -
e^{a\,\phi + 6\alpha\,\varphi}\, G_{\mu\nu}\Big)\,,\nn\\
&&\nabla^\nu\Big(e^{a\,\phi + 6\alpha\,\varphi}\, G_{\mu\nu}\Big)
=\ft12 e^{a\,\phi -4\alpha\, \varphi}\, G_{\mu\nu\theta}\,
F^{\nu\theta} + (3m_2 + 2m_1)\, e^{a\,\phi + 6\alpha\, \varphi}
\, G_{\mu\nu}\, \cA^\nu\,,\nn\\
&&\nabla^\nu\Big(e^{\ft12a\,\phi -2\alpha\,\varphi}\,F_{\mu\nu}^i
\Big) = -\ft12 e^{a\,\phi - 4\alpha\,\varphi}\, G_{\mu\nu\theta}\,
F^{i\,\nu\theta} - e^{a\,\phi + 6\alpha\,\varphi}\,
G_{\mu\nu}\, L^{i\,\nu}\nn\\
&&\qquad\qquad + (4m_2 + m_1)\, e^{\ft12a\,\phi - 2\alpha\,\varphi}
F_{\mu\nu}^i\,\cA^\nu - (4m_2 + m_1)\,e^{\ft12a\,\phi + 8\alpha\,
\varphi}\, L_\mu^i\,,\nn\\
&&\nabla^\mu\Big(e^{\ft12a\,\phi + 8\alpha\,\varphi}\, L_\mu^i\Big)
=\ft12 e^{\ft12a\,\phi -2\alpha\,\varphi}\, F_{\mu\nu}^i\,
\cF^{\mu\nu} + \ft12 e^{a\,\phi + 6\alpha\,\varphi}\,
G_{\mu\nu}\, F^{i\,\mu\nu}\nn\\
&&\qquad +(4m_2 + m_1)\, e^{\ft12a\,\phi + 8\alpha\,\varphi}\,
L_\mu^i\,\cA^\mu\,,\nn\\
&&\nabla^\nu\Big(e^{-10\alpha\,\varphi}\,\cF_{\mu\nu}\Big)
=\ft12 e^{a\,\phi - 4\alpha\,\varphi}\,G_{\mu\nu\theta}\,G^{\nu\theta}
-e^{\ft12a\,\phi -2\alpha\,\varphi}\,F_{\mu\nu}^i\,L^{i\,\nu}
+5m_2\, e^{-10\alpha\,\varphi} \cA^\nu\, \cF_{\mu\nu}\nn\\
&&\qquad\qquad+\sqrt{10}\,m_1\, (\del_\mu\phi - \sqrt{10}\,m_1\,\cA_\mu)
- 10m_2\, (\beta\,\del_\mu\varphi -m_2\, \cA_\mu)
\,,\nn\\
&&\square\phi = \ft1{3\sqrt{10}}\, e^{a\,\phi - 4\alpha\,\varphi}\,
G_\3^2 + \ft1{2\sqrt{10}}\, e^{\ft12a\,\phi -2\alpha\,\varphi}\,
(F_\2^i)^2 + \ft{1}{\sqrt{10}}\,e^{a\,\phi + 6\alpha\,\varphi}\,
G_\2^2\nn\\
&&\qquad + \ft{1}{\sqrt{10}}\,e^{\ft12a\,\phi + 8\alpha\,\varphi}\,
(L_\1^i)^2 + 5m_2\, \cA^\mu\, \del_\mu\phi -5\sqrt{10}\,
m_1\,m_2\, (\cA^2 + e^{10\alpha\,\varphi})\nn\\
&&\qquad + \sqrt{10}\,m_1\nabla_\mu \cA^\mu\,,\nn\\
&&-\beta\,\square\varphi = -\ft{1}{30}e^{a\,\phi - 4\alpha\,\varphi}
G_\3^2 - \ft1{20} e^{\ft12a\,\phi - 2\alpha\,\varphi}\, (F_\2^i)^2
+\ft3{20}e^{a\,\phi + 6\alpha\,\varphi}\, G_\2^2\nn\\
&&\qquad\quad -\ft14 e^{-10\alpha\,\varphi}\, \cF_\2^2
+\ft25e^{\ft12a\,\varphi + 8\alpha\,\varphi}\, (L_\1^i)^2
-5\beta\,m_2\, \cA^\mu\, \del_\mu\varphi\nn\\
&&\quad\qquad + 5m_2^2\, \cA^2 -
m_2\, \nabla_\mu\cA^\mu + 5m_1^2\, e^{10\alpha\,\varphi}\,,\nn\\
%%%
&&R_{\mu\nu} - \ft12g_{\mu\nu}\, R =
\ft12\del_\mu\varphi \del_\nu\varphi -\ft14 (\del\varphi)^2\,
g_{\mu\nu}+
\ft12\del_\mu\phi \del_\nu\phi -\ft14 (\del\phi)^2\,
g_{\mu\nu}\nn\\
&&\quad +\ft12 e^{-10\alpha\,\varphi}\,
(\cF_{\mu\theta}\,\cF_{\nu}{}^{\theta} - \ft14\, g_{\mu\nu}\,
\cF_\2^2) +\ft14 e^{a\,\phi-4\alpha\,\varphi}\,
(G_{\mu\theta\lambda}\,G_{\nu}{}^{\theta\lambda} - \ft16\, g_{\mu\nu}\,
G_\3^2)\nn\\
&&\quad +\ft12 e^{\ft12a\,\phi-2\alpha\,\varphi}\,
(F^i_{\mu\theta}\,F^i_{\nu}{}^{\theta} - \ft14\, g_{\mu\nu}\,
(F^i_\2)^2)
+\ft12 e^{a\,\phi + 6\alpha\,\varphi}\,
(G_{\mu\theta}\,G_{\nu}{}^{\theta} - \ft14\, g_{\mu\nu}\,
G_\2^2)\nn\\
&&\quad + \ft12 e^{\ft12a\,\phi + 8\alpha\,\varphi}\,(
L_\mu^i\,L_\nu^i - \ft12g_{\mu\nu}\, (L_\1^i)^2) -
5\alpha\,m_2\,
(\cA^\theta\,\del_\theta\varphi\, g_{\mu\nu} -
\cA_\mu\,\del_\nu\varphi-\cA_\nu\,\del_\mu\varphi)\nn\\
&&\quad +\ft12\sqrt{10}\, m_1\,
(\cA^\theta\,\del_\theta\phi\, g_{\mu\nu} -
\cA_\mu\,\del_\nu\phi-\cA_\nu\,\del_\mu\phi) -5(m_2^2-m_1^2)
\,\cA_\mu\,\cA_\nu\nn\\
&&\quad -\ft52m_2\, (\nabla_\mu\cA_\nu + \nabla_\nu\cA_\mu -
2\nabla_{\theta}\cA^\theta\, g_{\mu\nu}) -
5(2m_2^2 + \ft12 m_1^2)(\cA^2 + e^{10\alpha\,\varphi})\,
g_{\mu\nu}\,,
\eea
%%%%
where $a=4/\sqrt{10}$ and $\cF_\2=d\cA_\1$.  Clearly, these are the
bosonic equations of motion for the new $D=6$, ${\cal N}=(1,1)$ gauged
supergravity coupled to a vector multiplet.

       It would be interesting to obtain the pure $D=6$, ${\cal
N}=(1,1)$ gauged supergravity by truncating out the matter vector
multiplet.  To do this, we make an orthonormal rotation for the scalar
$\phi$ and $\varphi$, namely
%%%
\be
a\,\phi - 4\alpha\,\varphi = \sqrt2\,\phi_1\,,\qquad
4\alpha\,\phi + a\,\varphi=\sqrt2\,\phi_2\,.
\ee
%%%%
It now becomes clear that if we set $m_1=m_2\equiv m$, we can perform
the following consistent truncation
%%%%%
\be
\phi_2=0\,,\qquad \chi^i=0\,,\qquad
\cA_\1=-B_\1\equiv \ft1{\sqrt2}\,A_\1\,.
\ee
%%%%
The equations of motion for the remaining fields are then given by
%%%%%
\bea
&&\nabla^\theta\Big(e^{\sqrt2\,\phi_1}\,G_{\mu\nu\theta}\Big) =
\ft{5m}{\sqrt2}\,\Big(e^{\sqrt2\,\phi_1}\, G_{\mu\nu\theta}\,
A^\theta + e^{\ft1{\sqrt2}\,\phi_1}\, F_{\mu\nu}\Big)\nn\\
&&\nabla^\nu\Big(e^{\ft1{\sqrt2}\,\phi_1}\,F_{\mu\nu}\Big) =
-\ft12 e^{\sqrt2\,\phi_1}\, G_{\mu\nu\theta}\, F^{\nu\theta} +
\ft5{\sqrt2}m\, e^{\ft1{\sqrt2}\,\phi_1}\, F_{\mu\nu}\,A^{\nu}\nn\\
&&\nabla^\nu\Big(e^{\ft1{\sqrt2}\,\phi_1}\,F^i_{\mu\nu}\Big) =
-\ft12 e^{\sqrt2\,\phi_1}\, G_{\mu\nu\theta}\, F^{i\,\nu\theta} +
\ft5{\sqrt2}m\, e^{\ft1{\sqrt2}\,\phi_1}\, F^i_{\mu\nu}\,A^{\nu}\nn\\
&&\square \phi_1 = \ft1{6\sqrt2} e^{\sqrt2\,\phi_1}\, G_\3^2 +
\ft1{4\sqrt2}\, e^{\ft1{\sqrt2}\, \phi_1}\, (F_\2^2 + (F_\2^i)^2) +
\ft5{\sqrt2}\, m\, A^\mu\, \del_\mu\phi_1\nn\\
&&\qquad + \ft52 m\, \nabla_\mu\,A^\mu -
\ft{25}{\sqrt2}\, m^2\, (\ft12A^2 + e^{-\ft1{\sqrt2}\,\phi_1})\,,\nn\\
&&R_{\mu\nu} -\ft12g_{\mu\nu}\, R = \ft12 (\del_\mu\phi_1\, \del_\nu
\phi_1 - \ft12 (\del\phi_1)^2\, g_{\mu\nu}) +
\ft14 e^{\sqrt2\,\phi_1}\, (G_{\mu\theta\lambda}\,
G_{\nu}{}^{\theta\lambda} - \ft16 g_{\mu\nu}\, G_\3^2)\nn\\
&&\qquad +\ft12e^{\ft1{\sqrt2}\,\phi_1}\,(F_{\mu\theta}\,F_{\nu}{}^{\theta}
-\ft14 g_{\mu\nu}\, F_\2^2) + \ft12e^{\ft1{\sqrt2}\,\phi_1}\,
(F^i_{\mu\theta}\,F^i_{\nu}{}^{\theta}
-\ft14 g_{\mu\nu}\, (F^i_\2)^2)\nn\\
&&\qquad +\ft{5m}4\, (A^\theta\,\del_\theta\phi_1\, g_{\mu\nu} -
A_{\mu}\,\del_\nu \phi_1 - A_\nu\,\del_\mu\phi_1)\label{d6n11pure}\\
&&\qquad-\ft{5m}{2\sqrt2}\, (\nabla_\mu A_\nu +\nabla_\nu A_\mu -
2\nabla_\theta A^\theta\, g_{\mu\nu}) -
\ft{25}{2}m^2\, (\ft12 A^2 + e^{-\ft1{\sqrt2}\,\phi_1})\,g_{\mu\nu}
\,,\nn
\eea
%%%%
where $F_\2=dA_\1$, $F_\2^i=dA_\1^i$ and $G_\3 = dB_\2 + \ft12
F_\2\wedge A_\1 + \ft12 F_\2^i\wedge A_\1^i$.

    Thus we have obtained the full bosonic equations of motion for the
pure supergravity multiplet of the new $D=6$, ${\cal N}=(1,1)$ gauged
supergravity.  Note that all the four vectors are abelian, but with
the gauge symmetry of $A_\1$ broken owing to the higher-order
interactions.  This is different from the Romans $D=6$ gauged
supergravity where the four vectors are the $SU(2)\times U(1)$
Yang-Mills fields.  Furthermore, our new theory has a
positive-definite scalar potential $V=25m^2\, e^{-\ft1{\sqrt2}\,
\varphi}$.

        The supersymmetry of the theory is straightforward.  It was
shown in \cite{lav,berg} that the generalised dimensional reductions
are consistent with supersymmetry.  This can be seen from the fact
that the full $D=7$ equations of motion, including the fermions, are
invariant under the dilaton shifting symmetry and the homogeneous
scaling symmetry.

     As in the case discussed in \cite{lav}, the theory
(\ref{d6n11pure}) obtained from the generalised reduction does not
have a Lagrangian formalism.  This can be seen from the fact that if
there were a Lagrangian, it would from the Einstein equation of motion
in (\ref{d6n11pure}) have the term $m^2\, A^2$.  On the other hand,
the equation of motion for $A_\1$ indicates that such a term should
not exist.  At first sight, one might conclude that $A_\1$ is massive,
but the equation of motion for $A_\1$ clearly shows that it is a
massless field.

         A few remarks are needed at this stage.  Owing to the overall
$z$-dependent scaling factor in the ans\"atze (\ref{d7red1}) and
(\ref{d7red2}), the coordinate $z$ cannot be viewed as a circle
coordinate.  Thus the theory is not compactified.  To resolve such a
problem, it was proposed in \cite{hlw,chla} that one can modify the
original supergravity by introducing an auxiliary field associated
with the gauging of the scaling symmetry, which can be identified with
the reduction coordinate in the dimensional reduction.  The auxiliary
field always appear in the equations through a derivative in the
modified theory, and can therefore be defined as a circle coordinate
in the reduction.  Locally, this approach is the same as our
generalised circular reduction, but globally, the internal direction
is a circle instead of a real line.  In fact, if we consider in our
example the string frame, then there is no $z$-dependence in the
metric when $m_1=m_2$, and so $z$ can be viewed as a circular
coordinate at least from the metric point of view.  An alternative
approach is to introduce a delta function singularity \`a la
Randall-Sundrum.  We can then replace the prefactor in the metric
$e^{2m\, z}$ by $e^{-2m\, |z|}$.  By doing this, the volume of the
internal direction will be finite even though $z$ is a non-compact
coordinate.  Consequently, the gravity will be localised on the brane
located at $z=0$.  The exponential nature of the warp factor in the
conformal-frame metric implies that the effect of localisation is
strong with a mass gap.  It would be interesting to study further if
the delta function singularity in this procedure can be smoothed out.

   It is interesting to note that we can consistently truncate out the
$A_\1$ field in the pure bosonic equations of motion.  Then, the
reduction ansatz becomes
%%%
\bea
d\hat s^2_7 &=& e^{2m\, z}\, ( e^{-\ft1{5\sqrt2}\,\phi}\, ds_6^2 +
e^{\ft{2\sqrt2}{5}\, \phi}\, dz^2)\,,\nn\\
\hat B_\2 &=& B_\2\,,\qquad \hat A_\1^i = A_\1^i\,,\qquad
\hat \phi = \ft2{\sqrt5}\,\phi + \sqrt{10}\,m\,z\,.\label{d7red2}
\eea
%%%%%
Note that, if we consider the string frame, there is no $z$-dependence
in the metric.  The $z$-dependence in the reduction appears then only
in the dilaton.  The equations of motion for the remaining fields can
arise from the Lagrangian
%%%%%%%
\be
e^{-1}{\cal L} = R - \ft12(\del\phi)^2 - \ft1{12} e^{\sqrt2\,\phi}\,
G_\3^2 - \ft14 e^{\ft1{\sqrt2}\,\phi}\, (F_\2^i)^2 -
25 m^2\, e^{-\ft1{\sqrt2}\,\phi}\,,\label{d6lag}
\ee
%%%%%
where $G_\3 = dB_\2 + \ft12 F_\2^i\wedge A_\1^i$ and $F_\2^i=dA_\1^i$.
Intriguingly, this is precisely the Lagrangian for the bosonic sector
of the $D=6$, ${\cal N}=(1,0)$ gauged supergravity constructed in
\cite{szmodel}, coupled to two vector multiplets.\footnote{Note that
the bosonic sector of the minimal $D=6$, ${\cal N}=(1,0)$ gauged
supergravity consists of the metric, a dilaton, a tensor and a vector,
which, when ungauged, can be viewed as pure ${\cal N}=(1,0)$
supergravity coupled to a tensor and a vector multiplet} However,
unfortunately the truncation is not supersymmetric.  This can be seen
by setting the parameter $m=0$, in which case the truncation from
$N=(1,1)$ pure supergravity to $N=(1,0)$ requires setting to zero all
the four vector fields.

\section{M-theory interpretation}

     One intriguing feature of the $D=6$, ${\cal N}=(1,0)$ gauged
supergravity is that it admits a supersymmetric (Minkowski)$_4\times
S^2$ vacuum. In the new ${\cal N}=(1,1)$ theory, we also have such a
vacuum solution, given by
%%%%
\bea
ds^2 &=& dx^\mu\, dx^\nu\, \eta_{\mu\nu} +
\ft{1}{25m^2}\, d\Omega_2^2\,,\nn\\
F_\2 &=& \ft{\sqrt2}{5m}\, \Omega_\2\,,
\qquad \phi=0\,,\label{m4s2}
\eea
%%%%
where we have turned on one of the three vector field strengths
$F_\2^i$.  Lifting this solution back to $D=7$, it becomes
the near-horizon limit of a 3-brane supported by one of the vector
field strengths $\hat F_\2^i$.  To see this, let us start with the
3-brane, given by
%%%%
\bea
d\hat s_7^2 &=& H^{-\ft25}\, dx^\mu\, dx^\nu\, \eta_{\mu\nu} +
H^{\ft85}\,(dr^2 + r^2\,d\Omega_2^2)\,,\nn\\
\hat F_2 &=& \sqrt2\, Q\, \Omega_\2\,,\qquad e^{\phi} =
H^{-\ft2{\sqrt{10}}}\,, \label{d73brane}
\eea
%%%%%
where $H = 1 + Q/r$.  In the decoupling (or near-horizon) limit, we
have $H=Q/r$.  Taking the charge parameter $Q$ to be $Q=5m$, and
making a coordinate transformation $Q/r=e^{-5m\, z}$, the solution
(\ref{d73brane}) becomes
%%%%
\bea
d\hat s_7^2 &=& e^{2m\,z}\, (dx^\mu\, dx^\nu\,\eta_{\mu\nu} +
\ft{1}{25m^2}\, d\Omega_2^2 + dz^2)\,,\nn\\
\hat F_\2 &=& \ft{\sqrt2}{5m}\, \Omega_\2\,,\qquad
\hat \phi=\sqrt{10}\, m\, z\,.\label{d7sol}
\eea
%%%%
This fits exactly the reduction ansatz (\ref{d7red2}), giving rise to
precisely the lower dimensional solution (\ref{m4s2}).  It is worth
mentioning that the solution (\ref{d7sol}) can also be viewed as a
domain wall with a (Minkowski)$_4\times S^2$ world-volume.

    The Killing spinor for the 3-brane is given by
%%%%
\be
\hat \epsilon=e^{\ft12m\,z}\,\epsilon_0\,,\label{ks}
\ee
%%%%
where $\epsilon_0$ is a constant spinor satisfying $\hat
\Gamma_{z12}\, \epsilon_0=\epsilon_0$, where 1 and 2 are the labels
for the vielbein of $S^2$. Since the supersymmetry transformation rule
of the gravitino in $D=7$ has the form $\delta\hat\psi_\mu \rightarrow
\hat{\cal D}_\mu\hat\epsilon$, it follows that the generalised reduction
for the supersymmetry transformation parameter $\hat\epsilon$ is given
by $\hat\epsilon= e^{\ft12 m\,z}\, \epsilon_0$.  Thus the Killing
spinor of the 3-brane (\ref{ks}) fits precisely the reduction ansatz,
giving rise to a constant Killing spinor $\epsilon_0$ for the
(Minkowski)$_4\times S^2$ vacuum in $D=6$.

     We can further lift the solution back to $D=11$, where it becomes
the near-horizon structure of two intersecting M5-branes.  As in the
above, we start with the two intersecting M5-branes in $D=11$:
%%%%%
\bea
ds_{11}^2 &=& (H_1\,H_2)^{-1/3}\, \Big(dx^\mu\, dx^\nu\, \eta_{\mu\nu} +
H_2\, (dz_1^2 + dz_2^2) + H_1\, (dz_3^2 + dz_4^2)\nn\\
&& +
H_1\, H_2\, (dr^2 + r^2\, d\Omega_2^2)\Big)\,,\nn\\
F_\4 &=& (Q_1\, dz_3\wedge dz_4 + Q_2\, dz_1\wedge dz_2)\, \wedge
\Omega_\2\,,
\eea
%%%%%%
with $H_i=1 + Q_i/r$.  Setting $Q_1=Q_2=Q$, the solution in the
near-horizon limit becomes
%%%%
\bea
ds_{11}^2 &=& \rho^{2/3}\, (dx^\mu\, dx^\nu\, \eta_{\mu\nu} +
Q^2\,\fft{d\rho^2}{\rho^2} + Q^2\, d\Omega^2) +
\rho^{-\ft13}\, ds_4^2\,,\nn\\
F_\4 &=& Q\, J_\2\wedge\Omega_\2\,.\label{d11sol}
\eea
%%%%
Here we can replace the 4-torus $ds_4^2$ by a Ricci-flat K3 manifold,
and $J_\2$ is a self-dual harmonic 2-form in the K3.  It is
straightforward to see that the $D=11$ solution (\ref{d11sol}) becomes
(\ref{m4s2}) in $D=6$ by first reducing on the K3 manifold followed by
the generalised Kaluza-Klein reduction.

         It is interesting to note that only by taking the decoupling
or near-horizon limit does the brane solution fit the reduction
ansatz.  This is different from the usual Kaluza-Klein circle
reduction where the whole solution can be reduced instead of just the
near-horizon limit.  Thus the standard $S^1$ reduction can be viewed
as a special case of a DeWitt group-manifold reduction, whose
consistency is guaranteed, whilst the generalised Kaluza-Klein
reduction can be viewed as a special case of a Pauli sphere reduction,
where the consistency requires conspiracies.  (A discussion of the
terminology is contained in \cite{cglpsphere}.)

       Of course, the ${\cal N}=1$ supergravity in $D=7$ can also be
obtained from a $T^3$ reduction of the heterotic string theory, which
is S-dual to M-theory on K3.  The vector field strengths $F_\2^i$ in
the minimal $D=7$ supergravity come from setting equal the three
Kaluza-Klein and the three winding vectors.  It follows that the
3-brane in $D=7$ can be lifted to the $D=10$ heterotic theory as an
intersection of the heterotic 5-brane and Taub-NUT.

      In \cite{glps}, a more general class of dyonic strings were
obtained in the $D=6$, ${\cal N}=(1,0)$ gauged supergravity. Identical
solutions exist in our new gauged supergravity.  It is also
straightforward to lift these solutions to $D=7$ and hence further
back to $D=11$.  The $D=11$ structure is somewhat complicated,
supported by the 4-form field strength that carries three magnetic
charges and one electric charge; the solution is thus interpreted as
an intersection of three M5-branes and one M2-brane.

\section{Conclusions and discussions}

     In this paper we have performed generalised Kaluza-Klein
reduction of M-theory on K3$\times R$ and obtained a new $D=6$, ${\cal
N}=(1,1)$ gauged supergravity.  The theory differs from the Romans
gauged supergravity in that in the new theory all the four vector
fields are abelian, and it has in addition a positive definite scalar
potential.  We find that the theory admits a (Minkowski)$_4\times S^2$
vacuum solution, which can be embedded in the 3-brane (\ref{d7sol}) in
seven dimensions, which itself can be viewed as intersecting M5-branes
wrapping on a supersymmetric two-cycle of K3 in $D=11$.  Clearly, the
orders of the reductions of the 3-brane can be reversed, by performing
the $S^2$ reduction first, which gives rise to a $D=5$ domain wall,
with a (Minkowski)$_4$ world-volume.  We finally arrive at the
four-dimensional Minkowski spacetime by performing a brane-world
Kaluza-Klein reduction introduced in \cite{lpbrane}.  (See also,
\cite{clpbrane,dls,pps,lest,llpbrane}.)

       In fact, if we truncate out the Kaluza-Klein and winding
vectors, we expect that the above reduction can be performed on the
theory instead of just on a specific solution.  First, we expect that
there should be a consistent reduction of the minimal $D=7$
supergravity on $S^2$.  To see this, we can study the global symmetry
of the theory reduced on $T^2$.  If we for simplicity set two of the
three vector fields in $D=7$ to zero, then the resulting $D=5$ theory
has a global $O(2,3)$ T-duality symmetry, with the scalars
parameterising the coset $O(2,3)/(SO(2)\times SO(3))$.  Clearly, we
can gauge the $SO(3)$ maximal compact subgroup, which is exactly the
isometry group of $S^2$. This is indicative of a consistent $S^2$
reduction of the seven-dimensional theory \cite{clps3}.  The resulting
gauged $D=5$ supergravity will have a negative exponential-scalar
potential, which can support a domain wall solution. We can then
perform the brane-world Kaluza-Klein reduction to obtain a
four-dimensional ${\cal N}=1$ supergravity coupled to an $SU(2)$
vector multiplet and a scalar multiplet, exactly like the one obtained
from an $S^2$ reduction of the $D=6$ chiral gauged supergravity
\cite{gp}.  The fact that the brane-world Kaluza-Klein reduction can
give rise to Yang-Mills fields while keeping the cosmological constant
vanishing has been demonstrated in \cite{llpbrane}.

       It is also interesting to note that in our earlier approach,
the four dimensional theory arises first from the generalised
Kaluza-Klein reduction on $R$, and then the standard $S^2$ reduction,
in which case, the reduction makes use of a gauging of the homogeneous
scaling symmetry. If we instead perform the $S^2$ reduction first, and
then the brane-world reduction, it would appear that we do not need to
appeal to the homogeneous scaling symmetry.  Clearly, the two
approaches are equivalent. One feature in common is that in both
approaches, the reduction involves warp factors.  Thus our first
approach is nothing more than giving a symmetry interpretation of the
warp factor in the reduction ansatz.  In fact, the near-horizon
structure of the 3-brane given by (\ref{d7sol}) in $D=7$ can be viewed
as a domain wall written in the conformal frame, with the world-volume
being (Minkowski)$_4\times S^2$.  Thus the generalised dimensional
reduction can be viewed as a special case of the brane-world
reduction.

\section*{Acknowledgment}

       We are grateful to James Liu and Chris Pope for useful
discussions.

\end{document}